%% file: 2015-wraps.tex
\documentclass[10pt, conference, compsocconf]{IEEEtran}
\include{head}

\ifpdf
 \DeclareGraphicsExtensions{.pdf, .jpg, .tif}
\else
 \DeclareGraphicsExtensions{.pdf, .ps}
\fi

\newcommand{\B}[1]{\textbf{#1}\xspace}
\newcommand{\I}[1]{\textit{#1}\xspace}
\newcommand{\T}[1]{\texttt{#1}\xspace}

\newif\ifdraft
\draftfalse
\ifdraft
\usepackage{xcolor}
\newcommand{\jhanote}[1]{  {\textcolor{red}   { ***shantenu: #1 }}}
\newcommand{\amnote}[1]{   {\textcolor{blue}  { ***andre:    #1 }}}
\newcommand{\katznote}[1]{ {\textcolor{orange}{ ***dan:      #1 }}}
\newcommand{\note}[1]{     {\textcolor{green} { ***Note:     #1 }}}
\else
\newcommand{\onote}[1]{}
\newcommand{\terminology}[1]{}
\newcommand{\jhanote}[1]{}
\newcommand{\amnote}[1]{}
\newcommand{\katznote}[1]{}
\newcommand{\note}[1]{}
\fi

\newcommand{\synapse}{Synapse\xspace}
\newcommand{\Synapse}{Synapse\xspace}

\newcommand{\rp}{RADICAL-Pilot\xspace}
\newcommand{\enmd}{Ensemble-MD toolkit\xspace}

\DefineShortVerb{\|}

\DefineVerbatimEnvironment{mycode}{Verbatim}
{
  label=Code Example,
  fontsize=\small,
  frame=single,
  framerule=1pt,
  framesep=1em,
  numbers=left,
  gobble=2
}

\DefineVerbatimEnvironment{myio}{Verbatim}
{
  fontsize=\small,
  frame=lines,
  framerule=1pt,
  framesep=0.5em
}

\newenvironment{shortlist}{
  \begin{itemize}
}{
  \end{itemize}
}

\title{Synapse: Synthetic Application Profiler and Emulator} 
 
\author{
  \IEEEauthorblockN{
    Andre Merzky\IEEEauthorrefmark{1},
    Shantenu Jha\IEEEauthorrefmark{1}\\
  }
  \IEEEauthorblockA{
    \IEEEauthorrefmark{1}RADICAL Laboratory, Electric and Computer Engineering, Rutgers University, New Brunswick, NJ, USA
  }
}

\begin{document}

\maketitle
\begin{abstract}

  We introduce \synapse motivated by the needs to estimate and emulate workload
  execution characteristics on high-performance and distributed heterogeneous
  resources. \synapse has a platform independent application profiler, and the
  ability to emulate profiled workloads on a variety of heterogeneous
  resources. \synapse is used as a proxy application (or "representative
  application") for real workloads, with the added advantage that it can be
  tuned in different ways and at arbitrary levels of granularity in ways that
  are simply not possible using real applications.  Experiments show that
  automated profiling using \synapse represents application characteristics with
  high fidelity. Emulation using \synapse can reproduce the application
  behaviour in the original runtime environment, as well as reproducing properties
  when used in a different run-time environments.



\end{abstract}

%


%

\section{Introduction}
\label{sec:intro}



A large body of research in high-performance and distributed computing is
concerned with the design, implementation and optimization of tools, runtime
systems and services in support of scientific applications.  Scientific
applications are sophisticated entities, with complex development, deployment
and execution level requirements.  Furthermore, many scientific applications
have limited portability and scalability, thus restricting their direct use in
the development and optimization process of tools and services.


Some constraints and challenges associated with developing tools and services
can be addressed by the use of {\it synthetic applications}\footnote{also known
  as Application Skeletons, Representative Applications, and Artificial
  Applications} as proxies for the original application.  Synthetic applications
are viable proxies only when they can capture the essential characteristics of
the application being represented.  Synthetic applications should use a
relatively simple code base with minimal runtime requirements.  They should be
easy to deploy and easy to tune toward a specific use case and environment.


A tradeoff in the design and implementation of synthetic applications for use as
proxy applications is the need to be simple and general purpose on the one hand,
with the ability to emulate the behavior of application as accurately and as
with high fidelity as possible.  Achieving accuracy and high-fidelity is already
difficult if emulation is needed on multiple heterogeneous resources; the task
is made more difficult when those resources are different from the resource on
which the application(s) was profiled.

In response to these requirements and constraints, we have designed and
developed \synapse: a \I{SYNthetic Application Profiler and Emulator}.  \synapse
is primarily motivated by the need for automated and system-independent
application profiling in computational science, where the multitude and
generality of applications and platforms are the primary requirements, and not
cycle-level fidelity and very high-level precision.  Similarly, \synapse is
designed to provide uniform profiling capabilities across a range of application
types, tools and services.

\synapse acts as a simplified proxy application to circumvent the limitations
and complexity of the scientific application. For example, scientific
applications are not infinitely malleable, due to fixed and often discrete
physical sizes of input systems. They also typically have limited tunability, as
they can be modified only in discrete steps over a limited range of
values. \synapse can profile an application for given parameter values, but can
be tuned to emulate the same application for different parameter values.

\synapse is designed to ``profile once, emulate anywhere''.  \synapse determines
the application's resource consumption by running a sample based black-box
profiler on the application on \I{any} machine, and then replays the observed
consumption patterns on the target machine.  Thus, in emulation mode, \synapse
consumes precisely the same amount of resources (CPU, memory, storage, network)
as the original science application, without the need to profile the application
on the target machine.  Additionally, \synapse is able to estimate the resources
consumed, and the time-to-completion (TTC) on heterogeneous infrastructure
without necessarily having access to system-level capabilities.


While synapse is precise on \I{what} resource are consumed, \synapse is less
precise on exactly \I{how} those resources are consumed, i.e.,~in what
chunkiness, granularity and order.  We will discuss and verify that this
tradeoff is limited by \synapse's support for variable sample granularity and
its partial sample ordering approach.




Experiments validate the fundamental requirement that \synapse's
automated profiling captures the application characteristics with fidelity.
Experiments also show emulation using \synapse reproduces the application
characteristics in the original runtime environment as well as for different
resources and runtime systems.  While not designed to achieve the same accuracy
as other established approaches, experiments support the claim that \synapse's
emulation has sufficient fidelity and generality to make it a useful instrument
for the development of tools and services for computational science as well as
supporting computer science research.

This paper presents the initial design of \synapse and progress towards an
implementation that is robust and usable. In Section~\ref{sec:overview} we
outline three application and systems scenarios that have motivated the
development of \synapse. In Section~\ref{sec:arch}, we discuss the design and
architecture of \synapse, followed by the implementation in
Section~\ref{sec:impl}. Experiments are discussed in 
Section~\ref{sec:exp} followed future and related work.



\section{A Case for System-Independent Profiling and Emulation}
\label{sec:overview}

The development of tools for computational science, as well as for large-scale
computer science experiments need proxy applications that provide flexible and
tunable capabilities as well as being portable across resource types. We outline
three distinct use cases for proxy applications, each highlighting a different
requirement.

\paragraph{Abstractions and Middleware for Distributed Computing} 
In spite of significant progress in scientific distributed computing over the
past decade, there do not exist general purpose abstractions and middleware to
support the large-scale distributed execution of applications. Although many
specific solutions exist, they are customized to specific workloads and resource
types. In order to alleviate this shortcomings, as part of the DOE AIMES
project, we have designed the AIMES middleware for distributed
execution~\cite{ipdps-2016}. AIMES introduces the concept of {\it execution
  strategies} and we have demonstrated the use of AIMES abstractions and
middleware over different resource types. Initial progress was based upon using
static and some what simple workloads, such as bag-of-tasks of null workloads
(\texttt{sleep}). The challenges in generalizing the base capabilities to
different workloads on different resource types is more of an implementation and
deployment challenge than a conceptual challenge.  A proxy application that
could emulate actual workloads that would benefit from distributed execution
capabilities would play an important role in the validation and extension of
base AIMES abstractions and middleware. Proxy applications would have the
advantage of capturing important application properties without exposing the
complexity of running these applications on very distinct platforms.  


\paragraph{High-Performance Task-Parallel Computing} 
Traditionally high-performance computing (HPC) systems have been optimized to
support mostly monolithic workloads.  The workload of many important scientific
applications however, is comprised of spatially and temporally heterogeneous
tasks that are often dynamically inter-related~\cite{note4}. These workloads can
benefit from being executed at scale on HPC resources, but a tension exists
between their resource utilization requirements and the capabilities of HPC
system software and HPC usage policies.  In order to address this tension, we
have developed RADICAL-Pilot~\cite{review_radicalpilot_2015}, a scalable and
interoperable execution framework for task-level parallelism. \rp provides a
runtime system designed to support a large number of concurrent tasks with low
start-up overhead. It is agnostic to the specific properties of the executed
tasks, viz., many-node parallelism as well as single core tasks, short running
tasks versus long duration jobs.  Not surprisingly, there are many components
that need to be designed and parameterized in order to provide balanced
performance while being agnostic (as much as possible) to task properties. For
example, \rp's task execution component (the RP Agent) has to be engineered for
optimal resource utilization while maintaining the full generality.  

In practice, application workloads are not infinitely malleable, i.e., they can
only be modified in discrete steps over a limited range of values.  This often
necessitates the selection of new application workloads beyond certain ranges.
A proxy application could enable the design and test of \rp with a single
workload by providing the ability to tune ``application properties'' without
wholesale refactoring of the workload.

\paragraph{Toolkits for Computational Science} 
Many scientific applications in the field of molecular sciences, computational
biology \cite{preto2014fast}, astrophysics \cite{sirko2005initial}, weather
forecasting \cite{bauer2015quiet}, bioinformatics \cite{ martin2010rnnotator}
are increasingly reliant on ensemble-based methods to make scientific
progress. Ensemble-based applications vary in the degree of coupling and
dependency between the tasks, and heterogeneity across tasks. In spite of the
apparent simplicity of running ensemble-based applications, the scalable and
flexible execution of a large and collective set of tasks is non-trivial.  As a
consequence of complexity and many degrees-of-freedom, the challenges and the
growing importance and pervasiveness of ensemble-based applications, we designed
and implemented \enmd based upon a careful analysis of requirements of
ensemble-based applications. Similar to the previous two use-cases, a proxy
application would provide a lightweight and highly tunable workload so as to
simplify and design \enmd for general purpose workloads. In addition, a proxy
application would provide the ability to arbitrary vary the duration and number
of task instances between different stages of the application, as well as change
the coupling between tasks; this is an important characteristics of applications
used for advanced sampling~\cite{enmd}.



\section{\Synapse Scope and Architecture}
 \label{sec:arch}

 A finer-grained analysis of the aforementioned use cases, results in the
 following requirements on the profiling and emulation stages of \synapse.

 \subsection{Requirements on Application Profiling}
 \label{sec:prof_req}

  We state four requirements for correct profiling:

  \begin{shortlist}

    \item \B{P.1 Minimal Self-Interference:} the act of profiling does
        not influence the results of the profiling; 

    \item \B{P.2 Low Overhead:} the act of profiling does not
        influence the runtime behavior of the profiled application;

    \item \B{P.3 Black-box Approach:} the act of profiling does not
        require any changes in application code, and minimal,
        non-intrusive changes in application runtime environment;

    \item \B{P.4 Consistency:} repeated profiling of the same
        application, in the same environment, yields consistent
        results, and the results of profiling are usable to reproduce
        (emulate) the application's runtime behavior;
        
  \end{shortlist}

  We believe these requirements are both necessary and sufficient, given the
  scope and motivation of this work.  We do not directly list 'quantitative
  correctness' as a requirement, i.e., we do not consider it strictly necessary
  for profiling to measure the \I{exact} number of Bytes and FLOPs etc, but
  rather only require that the measured metrics are reliable and consistent, and
  allow truthful application emulation.  Having said that, we do believe that
  the metrics that we will discuss in Section~\ref{sec:metrics} are relevant in
  the sense that they reflect executed resource level operations. Some caveats
  that are discussed later.

 \subsection{Requirements on Application Emulation}
 \label{sec:emu_req}

 Requirements for the application emulation as motivated are:

  \begin{shortlist}

    \item \B{E.1 Fidelity:} application emulation must exhibit the
        same runtime characteristics as the execution of the actual
        application.  Amongst others, we specifically expect emulation
        TTC to correspond to application TTC. 

    \item \B{E.2 Portability:} the application can be emulated on
        resources other than the one used for profiling.

  \end{shortlist}


 \subsection{Synapse Architecture}

 \Synapse is a research prototype used in support of other research
 projects.  As such it is subject to frequent changes in in target use cases,
 requested features and system to be supported.  We chose an architecture which
 is modular for both the profiling as the emulation part of Synapse, but is
 otherwise thin and lightweight.  The implementation in Python caters toward
 portability and usability, with some caveats which are discussed in the
 implementation description in Section~\ref{sec:impl}.

  \begin{figure}[h]
   \includegraphics[width=0.50\textwidth]{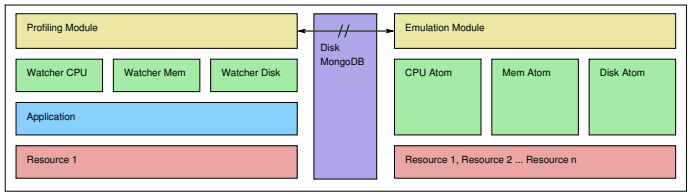}
   \caption{\textbf{\Synapse Architecture:} The profiling component
       manages a set of watcher plugins, which observe the runtime
       behavior of an application execution.  The emulation component
       interprets the resulting profiles to control a set of
       'emulation atoms' which behave similarly, and can be run on any
       resource.  Profiles are stored on disk or in a MongoDB
       instance.
     \label{fig:arch}
   }
  \end{figure}

  Figure~\ref{fig:arch} shows the resulting \synapse architecture.
  The modularity of the \synapse profiling is provided by extensible
  and exchangeable \I{Watcher} plugins which profile for a very
  specific resource type; the counterpart on the \synapse emulation
  are pluggable emulation \I{Atoms} which can emulate the consumption
  of the same resource types as profiled by the Watchers.  Profiles
  are stored in local |json| files, but alternatively \synapse can
  employ a MongoDB database for storing collected profiles, and also
  for retrieving profiles for later emulation runs.

%
\section{Implementation of \Synapse}
\label{sec:impl}

 \synapse is implemented as a Python module\footnote{Some well
     contained parts of synapse are written in C and assembly, as
     discussed later on.} that primarily provides two methods:

 \begin{myio}
 radical.synapse.profile(command, tags=None)
 radical.synapse.emulate(command, tags=None)
 \end{myio}

 \noindent where |command| is either a shell command line or a Python
 callable (which is then spawned in its own Python shell).
 
 The |profile| method profiles the specified command and stores the
 result in a MongoDB database, which is indexed by command itself and
 by an optional set of user supplied tags.  Repeated profile runs of
 the same command/tag combination will let \synapse collect multiple
 profiles for statistical analysis, to estimate means and standard
 deviation of resource consumption metrics.  The tags are used to
 differentiate application instances that are not distinguishable by
 their command line alone, but that are expected to result in
 different execution profiles.  For example, tags can be used to flag
 different semantic content of application control and configuration
 files, or to distinguish between different application
 configurations.

 The |emulate| method utilizes a set of emulation \I{atoms}, which are
 very fine grained and tunable software elements which consume one
 specific type of resource.  Synapse uses the command and tag
 combination specified on the |emulate| call to search the database
 for any available profile.  Once a profile is found, \synapse will
 retrieve the set of profile samples and will feed them to the
 emulation atoms in the order in which they have been collected.  That
 sample ordering is an essential element for the fidelity of the
 emulation, as will be discussed in detail in
 Section~\ref{sec:sampling}.  Some very light-weight profiling is also
 applied during \synapse's own emulation run, to verify that the
 resources are consumed as expected.

 \Synapse also provides a set of command line tools which are
 essentially wrappers around certain configurations and combinations
 of the |profile| and |emulate| module calls.

 \subsection{Implementation of \Synapse Profiling}
 
 The \synapse profiler relies on several system utilities.  Amongst
 others, it uses the |perf-stat| utility to inspect CPU activity, the
 |/proc/| filesystem to read system counters on memory and disk I/O,
 and the POSIX |rusage| call to obtain runtime process information.

 The different information providers are implemented as plugins,
 \synapse is thus extensible with additional profiling metrics (see
 discussion of future work in Section~\ref{sec:future}). Those
 plugins are structured as follows:

 \begin{myio}
 class WatcherClass(rsw.WatcherBase):
     def __init__(self, pid):
         ...
     def _pre_process(self, config): 
         ...
     def _sample(self): 
         ...
     def _post_process(self): 
         ...
     def _finalize(self): 
         ...
 \end{myio}

 Pre- and Post-Process set up and tear down any profiling environment
 for that watcher.  The |_sample| method is invoked at regular
 intervals by the main \synapse profiling loop.  In the |_finalize|
 method, the plugin has access to the raw profiling results of other
 watchers, in order to perform some further post processing.  While this
 creates some dependencies between plugins, it prevents the
 duplication of measurements (such as overall runtime).

 Each watcher plugin runs in its own thread:

 \begin{myio}
 def run(self):
     self._pre_process(self._config)

     while not self._terminate.is_set():
         now = timestamp()
         self._sample(now)
         time.sleep(self._sample_rate)

     self._post_process()
 \end{myio}

 Once \synapse spawns the application process, it communicates the
 process PID to the watcher threads, and they begin monitoring the
 process.  There is a small delay between process spawning and start
 of profiling but the process itself is wrapped into the POSIX tool
 |time -v|, which allows us to correct some of the effects of that
 offset\footnote{Other effects are found to be too small to matter.
 The first watcher sample is usually collected at around $0.005$
 seconds after startup.}.

 Profile data are collected as time series.  The timestamps of the
 different watchers are not synchronized, and can drift relative to
 each other over time.  We found this preferable to an increased
 profiling overhead due to synchronization.  The individual time
 series are combined during postprocessing and pushed into a MongoDB,
 or are written to disk.

 The sample rate is globally controlled via an environment variable,
 and is uniform over all watchers.  The highest sample rate is $10$,
 i.e., \synapse can at most gather one sample every $100 ms$.  That limit
 coincides with the limit of |perf stat|.  There is no lower bound to
 the sampling rate.  Section~\ref{sec:exp} discusses the impact of
 different sampling rates on profiling overhead, profiling accuracy,
 and emulation fidelity.

 \subsection{Implementation of \Synapse Emulation}

 At its core, the \synapse emulation framework consists of a set of
 small, self-contained C-codes (\I{\synapse Atoms}) that consume one
 specific resources type.  Currently, compute, memory, storage and
 network atoms have been implemented.

 The compute atom contains a loop of assembly code that efficiently
 performs a matrix multiplication.  The loop's efficiency represents
 the maximum efficiency \synapse can emulate, which seems on par with
 the various application codes we have profiled so far.  The
 efficiency of the assembly loop can be artificially lowered toward
 the target emulation efficiency by reducing the loop invocation
 frequency.  In all cases we make sure that the matrix size is small
 enough to fit fully into the CPU caches.

 The memory and storage atoms are relatively simple C codes that
 perform the respective canonical |libc| operations (|malloc|, |free|,
 |read| and |write|).  Those operations use buffer sizes that can be
 tuned, but are ultimately independent of the buffer sizes used in the
 actual application.  This introduces potential discrepancies compared
 to the emulated application, since system performance directly
 depends on the buffer size of I/O operations.  Our assumption is that
 application codes are generally aware of this, and attempt to use
 large block sizes where possible, and that small reads/writes are
 dominantly served by disk caches, and have thus
 relatively small impact on the overall performance.  In
 Section~\ref{sec:exp} we support that assumption with experiments,
 but acknowledge that it is likely to break (to a varying degree) for
 certain types of applications that are bound by specific I/O
 patterns.
 
 The \synapse profiler features an experimental watcher plugin that
 can, in principle, infer block sizes of disk I/O operations with
 |blktrace|.  We consider using this data in \synapse emulation when
 applications require that granularity to be future work (see
 Section~\ref{sec:future}). 

 The \synapse emulation atoms are driven by a global loop which feeds
 sequences of profile samples to the atoms for emulation.  The sample
 granularity is the same as used for profiling: the profiling sampling
 rate thus not only determines the accuracy of the profiling itself,
 but also influences the emulation fidelity.  All atoms run in
 separate processes; resources are thus utilized concurrently.  That
 may or may not reflect what the application code implements.  While
 the profiler does gather information about the number of used
 application threads and processes, that information at the moment is
 not used in the \synapse emulation phase.
 %

 \subsection{Profiling Metrics}
 \label{sec:metrics}

 Three main types of resources are currently profiled:
 compute (CPU), storage (disk), and memory.  \synapse measures several
 metrics for each of those, as listed in Table~\ref{tab:metrics}.
 Additionally, \synapse records several types of system information,
 such as number and type of CPU cores, available memory, and system
 load.  Some of those are used to compute derived metrics, for
 example, the CPU type and clock speed determines the maximum number
 of operations per second, which when combined with the observed
 number of used and stalled instructions cycles yields CPU efficiency
 and utilization.

 \synapse is able to force an artificial CPU, disk and memory load onto
 the system while emulating an application, thus emulating the
 application execution in a stressed environment.  We do not currently
 measure the disk and memory stress on the system, so these load
 factors have to be specified manually, and are currently used to
 confirm \synapse's viability on stressed systems.  Artificial load has
 not been used in the experiments presented in this paper, and is thus
 not discussed any further.

 Several listed metrics, such as CPU efficiency and utilization, are
 marked as \I{derived}: they are not directly reported by the system,
 but are calculated from other, primary metrics.  We use the following
 formula to compute CPU efficiency:

 \begin{eqnarray*} \mathit{efficiency} &=& cycles_{used} /
     cycles_{spent}\\ &=& cycles_{used} / (cycles_{used} +
     cycles_{wasted}) \end{eqnarray*}\\[-1em]

 \noindent
 We interpret the |'cycles'| reported by |perf stat| as
 $cycles_{used}$, and |'cycles_stalled_frontend'| |+|
 |'cycles_stalled_backend'| as $cycles_{wasted}$, the latter in the
 sense that they are counted toward the application execution, but did
 not contribute to its progression.  This is not a canonical
 definition for CPU efficiency, because the wasted cycles can
 potentially be counted twice (once for frontend and once for
 backend), or can overlap with used cycles (the backend can be busy
 while the frontend stalls).  However, the metric makes semantic sense
 in that it considers used cycles to contribute to higher efficiency,
 and any stalling to lower efficiency, which reflects the intuitive
 interpretation of those values in terms of code efficiency.

 Similarly, we compute CPU utilization as:

 \begin{eqnarray*}
     utilization &=& cycles_{used} / cycles_{max}
 \end{eqnarray*}\\[-1em]

 \noindent
 where $cycles_{max}$ is derived from the maximum possible number of
 cycles, which is determined by the CPU architecture and clock speed.
 \synapse does not sample the CPU clock speed (modern CPUs can adapt
 clock speed to load to preserve energy), and we do not take any
 background CPU activity (by the system or other applications) into
 account.  The derived utilization is still a useful metric to
 interpret in that it exposes the expected monotonic behavior toward
 faster/slower execution, but it is not comparable to similar metrics
 derived by other software.

 \begin{table}[t]
  \begin{center}
   \begin{tabular}{llcccc}
    \toprule
    \B{Resource} & \B{Metric}              & \B{Tot.} & \B{Samp.} & \B{Der.} & \B{Emul.}\\\midrule
    \B{System }  & number of cores         &  \T{ + } &   \T{ - } &  \T{ - } &  \T{ - } \\
                 & max CPU frequency       &  \T{ + } &   \T{ - } &  \T{ - } &  \T{ - } \\
                 & total memory            &  \T{ + } &   \T{ - } &  \T{ - } &  \T{ - } \\
                 & runtime                 &  \T{ + } &   \T{ + } &  \T{ - } &  \T{ - } \\
                 & system load (CPU)       &  \T{ + } &   \T{ - } &  \T{ - } &  \T{ + } \\
                 & system load (disk)      &  \T{ - } &   \T{ - } &  \T{ - } &  \T{ + } \\
                 & system load (memory)    &  \T{ - } &   \T{ - } &  \T{ - } &  \T{ + } \\\midrule
    \B{Compute}  & CPU instructions        &  \T{ + } &   \T{ + } &  \T{ - } &  \T{ + } \\
                 & cycles used             &  \T{ + } &   \T{ + } &  \T{ - } &  \T{ - } \\
                 & cycles stalled backend  &  \T{ + } &   \T{ + } &  \T{ - } &  \T{ - } \\
                 & cycles stalled frontend &  \T{ + } &   \T{ + } &  \T{ - } &  \T{ - } \\
                 & efficiency              &  \T{ + } &   \T{ + } &  \T{ + } &  \T{(+)} \\
                 & utilization             &  \T{ + } &   \T{ + } &  \T{ + } &  \T{ - } \\
                 & FLOPs                   &  \T{ + } &   \T{ + } &  \T{ + } &  \T{ - } \\
                 & FLOP/s                  &  \T{ + } &   \T{ + } &  \T{ + } &  \T{ - } \\
                 & threads                 &  \T{ + } &   \T{ - } &  \T{ - } &  \T{ + } \\\midrule
    \B{Storage}  & bytes read              &  \T{ + } &   \T{ + } &  \T{ - } &  \T{ + } \\
                 & bytes written           &  \T{ + } &   \T{ + } &  \T{ - } &  \T{ + } \\
                 & block size read         &  \T{ - } &   \T{(-)} &  \T{ - } &  \T{(-)} \\
                 & block size write        &  \T{ - } &   \T{(-)} &  \T{ - } &  \T{(-)} \\\midrule
    \B{Memory}   & bytes peak              &  \T{ + } &   \T{ + } &  \T{ - } &  \T{ - } \\
                 & bytes resident size     &  \T{ + } &   \T{ + } &  \T{ - } &  \T{ - } \\
                 & bytes allocated         &  \T{ + } &   \T{ + } &  \T{ + } &  \T{ + } \\
                 & bytes freed             &  \T{ + } &   \T{ + } &  \T{ + } &  \T{ + } \\
                 & block size alloc        &  \T{ - } &   \T{(-)} &  \T{ - } &  \T{(-)} \\
                 & block size free         &  \T{ - } &   \T{(-)} &  \T{ - } &  \T{(-)} \\\midrule
    \B{Network}  & connection endpoint     &  \T{(-)} &   \T{(-)} &  \T{ - } &  \T{(+)} \\
                 & bytes read              &  \T{(-)} &   \T{(-)} &  \T{ - } &  \T{(+)} \\
                 & bytes written           &  \T{(-)} &   \T{(-)} &  \T{ - } &  \T{(+)} \\
                 & block size read         &  \T{ - } &   \T{(-)} &  \T{ - } &  \T{(-)} \\
                 & block size write        &  \T{ - } &   \T{(-)} &  \T{ - } &  \T{(-)} \\\midrule
   \end{tabular}
   \caption{
       \B{List of \synapse metrics and their usage}\newline
       \B{Sampl.:} sampled over time;
       \B{Der.:} derived from other metrics;\newline
       \B{Tot.:} integrated total over runtime;
       \B{Emul.:} used in emulation;\newline
       \B{\T{(+)}:} partial;
       \B{\T{(-)}:} planned.
       \label{tab:metrics}
   }
  \end{center}
 \end{table}

 Table~\ref{tab:metrics} includes several metrics that are currently
 planned or only partially implemented.  Specifically, it lists
 network interactions, which \synapse can to some extent emulate, but
 which are not yet meaningfully profiled.  
 %
 %
 CPU efficiency is listed as `partially supported' for emulation:
 \synapse is able to tune the CPU load toward a certain efficiency
 value, but that tuning is currently manual (the experiments presented
 in the paper use the default values for all tunable settings).

 \subsection{The Effects of Sampling}
 \label{sec:sampling}

\begin{figure}[b]
   \centering
   \includegraphics[width=0.50\textwidth]{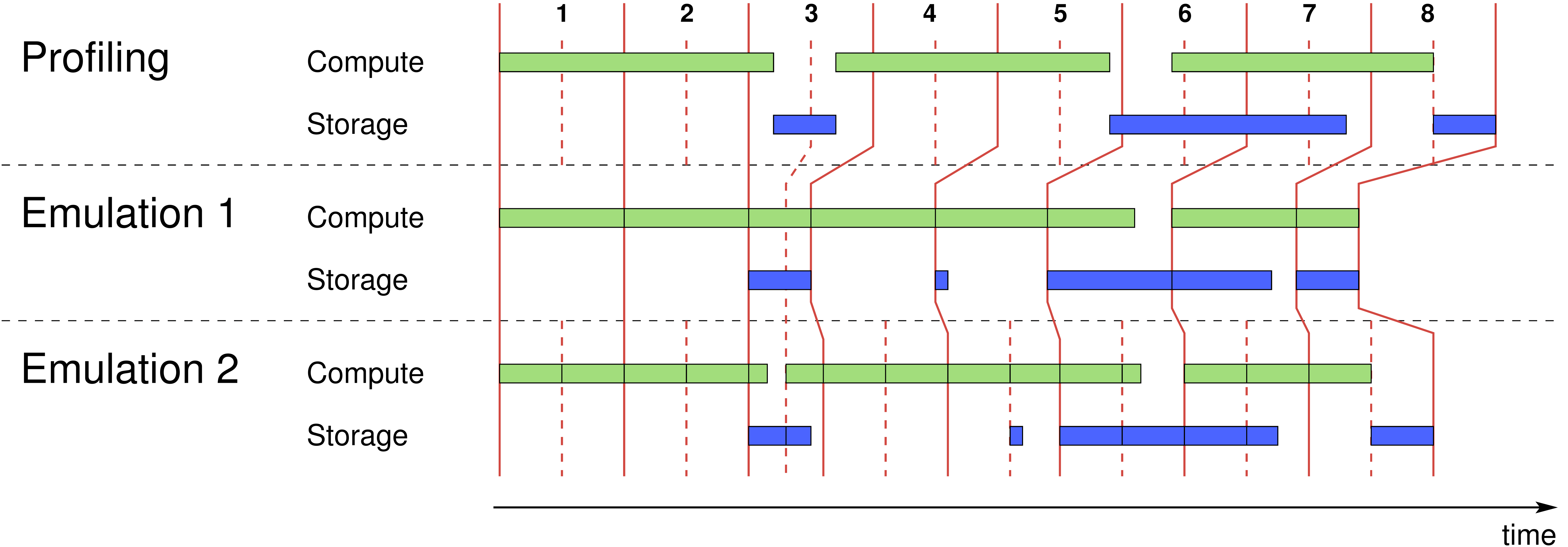}
   \caption{\textbf{Sampling Effects:} the profiling (top) shows
     a mix of serial and concurrent CPU (green) and disk (blue)
     utilization.  Solid red lines represent profiling sample
     boundaries; broken red lines represent sample boundaries at
     doubled sampling rate.  Emulations 1 and 2 (middle, bottom)
     replay different sample types (compute, storage) concurrently,
     thus removing some of the serialization of the original resource
     consumption (see sample numbers 3, 8).  Emulation 2 replays the
     higher sampling rate, thereby reducing that effect by partially
     re-introducing the serialization of the original resource
     consumption (see again samples 3, 8).
     \label{fig:sampling_1} 
     } 
  \end{figure}

  The effects of sampling are illustrated in
  Figure~\ref{fig:sampling_1}.  Profiling metrics are gathered at
  (roughly) equidistant points in time, for different types of
  resources.  Emulation follows the same clustering, but \I{disregards
  all timing information}.  It is after all not the purpose of
  emulation to reproduce the exact same timings, but to consume the
  exact same resources.  We will discuss several detail of this figure
  below.

  Figure~\ref{fig:sampling_1} illustrates that profiled resource
  consumptions may or may not fill a complete sampling period.  Where
  one specific resource interaction dominates overall application
  performance for that sample, one can expect that that type does fill
  a sampling period (e.g., samples 1 \& 6).  In the general case,
  a sampling period will capture several full or partial resource
  consumption types which may or may not occur concurrently.

  What resource consumption operation is accounted for in what sample
  depends on a multitude of parameters: applications will often employ
  techniques to hide I/O latency, such as caching or asynchronous
  operations, and the operating system itself uses latency hiding
  (caches, read-ahead, branch prediction etc).  In those cases, actual
  system activity can occur before or after the application code
  requests it.
  
  During \synapse's emulation, all resource consumptions for
  a specific sample are started immediately and concurrently upon
  starting that sample, without any ordering in between resource
  types.  Emulation samples end when the last resource consumptions is
  completed for that sample, and then the emulation for the next
  sample is started (see samples 3 \& 4 in
  Figure~\ref{fig:sampling_1}).  Resource consumptions that are not
  concurrent in the application \I{are} concurrent in the emulation
  (see samples 3 and 8), thus yielding potential emulation speedup.
  Smaller samples reduces that effect (see Emulation 2 in the figure,
  alternative samples 3 \& 8).  
  

  In many cases, one type of resource consumption is a semantic
  requirement for another type.  For example, an application needs to
  read data from a disk before being able to compute on those data; it
  needs to allocate memory before reading data from disk into that memory;
  it needs to perform computation before being able to write results
  to a disk; etc.  The code-agnostic sampling approach used by
  \synapse does not allow to directly detect such dependencies --
  parts of those dependency information are, however, implicitly
  captured: operations observed in a sample at time $t_n$ can only
  depend on resource consumption at samples from $t_{n-1}$ or
  earlier.  By
  ensuring that the emulation respects sampling order across resource
  types, \synapse will implicitly play back the dependencies thus
  captured.

 \begin{figure}[t]
   \centering
   \includegraphics[width=0.50\textwidth]{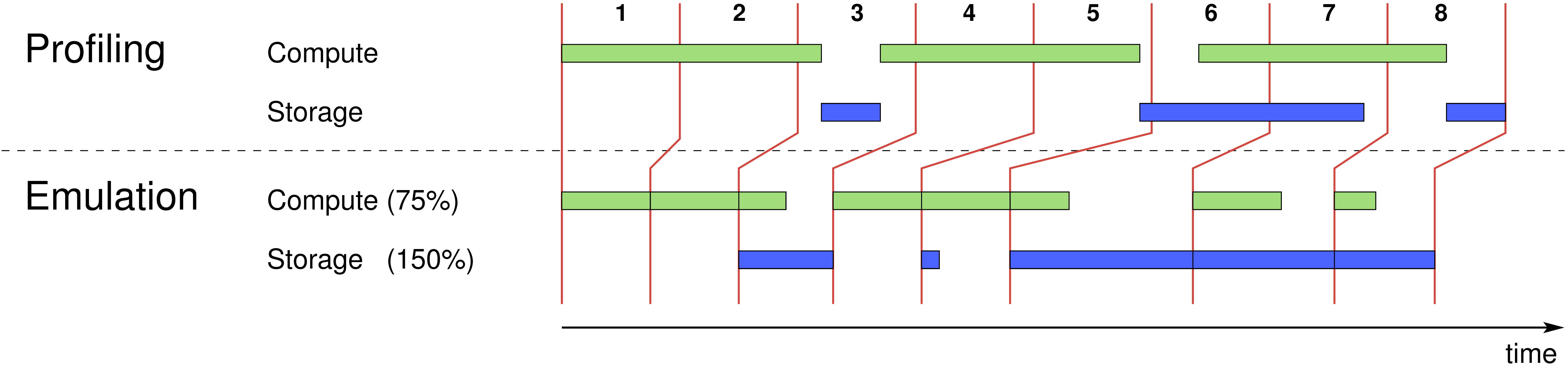}
   \caption{\textbf{Sample Portability:} the same profile samples as
       in figure~\ref{fig:sampling_1} is shown on top.  The bottom
       shows the emulation on a resource with different performance
       (CPU is $25\%$ faster, disk is $50\%$ slower).  The dominating
       resource type switches for some samples (3, 6, 8), but the
       overall activity ordering is preserved.
     \label{fig:sampling_2} 
   } 
   \end{figure}

  \synapse profile samples are designed to be portable, i.e., they can
  be used to emulate the application on resources other than the
  profiling resource.  Figure~\ref{fig:sampling_2} illustrates that
  the implicit dependencies captured in the sampling order preserves
  the order of the original application activities even on machines
  with very different performance characteristics, reflecting that the
  dominating contribution to the application's TTC can differ per
  machine.  In the figure, we see that the emulation is performed on
  a machine with faster CPU, but slower disk.  While sample 7 on the
  originally profile machine is, for example, dominated by the
  application's CPU utilization, the same sample is storage I/O
  dominated during emulation.

 \subsection{Scope and Limitations}
 \label{sec:limits}

 Sections~\ref{sec:intro} and~\ref{sec:overview} motivated the scope
 for which \synapse was defined.  This subsection makes this scope
 more specific; we list the set of conditions under which
 \synapse is expected to operate, or under which it is not.

 \subsubsection{Application Semantics}  \synapse watches
 \I{application behavior} -- it explicitly does not inspect the
 application at the code or system call level, and thus has no
 knowledge whatsoever of application semantics.  This limits the
 applicability of \synapse in some contexts.  For example, the POSIX
 system call |sleep(3)| will consume a very small number of flops, but
 will show significant contributions to TTC.  An inspection on
 different layers (code, libc call, OS signals etc.) could reveal that
 behavior, but that is considered out of scope for \synapse.

 \subsubsection{Resource Details}  A side effect of not inspecting the
 application on code or system call level is that \synapse does not
 distinguish which exact system resource is used: for example, it
 finds that the application wrote a certain number of bytes to disk,
 but does not infer what file system the data have been written to (we
 currently assume that all I/O operations go to |/tmp/|).  That can
 though significantly impact application performance, specifically for
 HPC resources that often feature shared file systems.We consider the
 use of |blktrace| for I/O profiling on the level of individual block
 device operations.

 \subsubsection{Application Granularity}  A similar side effect of
 external, sampled measurements is that application activities are not
 resolved beyond a certain granularity.  For example, \synapse can
 measure the number of bytes written to disk in a certain period of
 time, but actual I/O performance can vary significantly depending on
 \I{how exactly} those I/O operations are executed: a large number of
 small, scattered I/O operations will often be much slower than a
 small number of large I/O operations.  \synapse does not distinguish
 those cases: it assumes a static block size for the emulation of
 I/O (that block size can be manually tuned, but was left at the
 default value for the experiments in this paper).

 \subsubsection{Application Optimization}  Different resources may
 provide different means to optimize application codes, via compiler
 flags, optimized system libraries, specific hardware etc.  \synapse's profiling on one
 system cannot take optimization on another system into account, when
 those optimizations map to different resource consumption patterns,
 such as GPU acceleration which is available on the target host but
 was not used on the profiling host.  Profile portability is thus
 limited to resources with fundamentally similar architectures.
 The experiments in this paper were done with application code that
 was compiled with default settings for each resource, and that uses
 optimized system libraries where available.

 \subsubsection{Multithreading}  Application performance varies
 significantly with the number of threads employed to perform the
 necessary operations.  While \synapse does record the number of
 application threads, it does not distinguish what operation
 originates in what thread, nor does it use that information during
 emulation (all emulation is multi-processed).  The sampling based
 approach provides some mitigation to this, as it infers dependencies
 between data and compute operations, as discussed in
 Section~\ref{sec:impl}.  That inference can be wrong though, and the
 recorded order of events can be a coincidence.  In that case, the
 sampling based emulation will introduce too many synchronizations,
 and emulation will be slower than the actual application.  This specifically can
  happen for target resources where resource types have
 very different performance (e.g., a much faster disks).  Whenever an
 application is bound by a \I{single} resource type, that
 reordering effect will not apply.

 \subsubsection{Multiprocessing}  \synapse's profiling is process
 based -- it targets single-process applications.  \synapse does not
 attempt to detect the spawning of additional application processes.
 This could in principle be added (|/proc/| contains the required
 information), but support is not planned at this point.

 \subsubsection{IPC}  \synapse will not detect interprocess
 communication, neither between processes within the same OS, nor any
 communication over the network.  Specifically it is not able to
 handle application level threading locks, semaphore locks, etc.  We
 plan to at least add profiling of MPI communication at some point (see
 Section~\ref{sec:future} on future work), most likely by utilizing
 one of the many existing MPI profilers.

 \subsubsection{Overhead}  The processes of profiling and emulation
 consume certain amounts of resources.  \synapse manages though to
 keep those overheads very small (see experiments in
 Section~\ref{sec:exp}).
 The profiler's startup-time is
 constant and in the order of $<O(1)$ seconds.  Concurrent to the
 application, the profiler consumes a part of another CPU core (if
 available), and about 150~MB of memory.  A very high sampling rate
 can increase the overall memory footprint.  Writing the data to
 the database requires some time, depending on network latency and
 total number of samples.
 \synapse emulation has a similar overhead (fetching the samples from
 the DB into memory, a loop that feeds the \synapse atoms).  

 The emulation additionally shows some memory overhead.  This is
 partially owed to the fact that multiple python instances are
 spawned, and Python is often more memory heavy than the (compiled)
 application codes under investigation.  That memory overhead though
 is not large enough to significantly influence the measured TTCs, but
 it does show up in the profiles of emulation runs.

 Profiling will only terminate when full sample periods have passed,
 which can thus delay the completion of the profiling process to up to
 one additional sampling period.  That is only relevant for very low
 sampling rates.

 \subsubsection{DB limitations}  MongoDB has a 16~MB limit on the size of
 a single document.  This limits the total number of data samples
 \synapse supports to about 250,000.  This limitation can be lifted by
 changing to a different data model or storage backend.  File based
 storage of profiles is available.

%
\section{Experimental Results and Discussion} 
\label{sec:exp}

Our experiments are designed to investigate the viability of \synapse's approach
as a tool that, (i) automatically derives application profiles, and (ii)
implements synthetic application components which can emulate the profiled
applications.  The experiments demonstrate the fidelity of \synapse's profiling
and emulation for a specific scientific application under a range of conditions,
on a range of resources. Experiments are designed to support the requirements
listed in Sections~\ref{sec:prof_req} and~\ref{sec:emu_req} and cover the
following steps:


 \begin{itemize}
  
  \item Use \synapse to profile an application over a range of
      application parameters, with different sampling rates.

      \B{Purpose:} determine the profiling overhead versus
      non-profiled execution (experiment 1: P.1/P.2); show how the
      consistency of profiling results depends on sampling rate
      (experiment 2: P.3).

  \item Use \synapse to emulate the same application over the same
      range of application parameters, measuring TTC.

      \B{Purpose:} show the fidelity of the profiling results to capture
      relevant application characteristics; determine emulation precision in
      computing application TTC compared to actual execution (experiment 3: P.4,
      E.1).

  \item Use \synapse to emulate the same application on different
      resources, measuring TTC.
      
      \B{Purpose:} support the claim that \synapse profiling metrics are system
      independent; determine emulation precision in computing application TTC,
      on resources that are distinct from those used for profiling (experiment 4: P.4, E.2).

 \end{itemize}

 The application used for all experiments is Gromacs~\cite{pronk2013gromacs}. It
 is an application used for Molecular Dynamics (MD) simulations, in particular
 for biomolecular simulations. Gromacs is used by
 thousands of scientists, including multiple collaborators of the
 authors. 


 We configured applications with a varying number of
 iteration steps, ranging from $10^4$ to $10^7$.  The number of steps influences
 both CPU consumption and disk output, but leaves disk input and memory
 consumption constant.
 
 Due to space constraints, we do not present other experiments, but note that
 variations in other parameters also yield variations in the application's
 resource consumptions, which can also be captured by \synapse profiling and
 represented by \synapse emulation.  Ultimately, \synapse will not care
 what parameter changes cause the variation, as the application execution is
 considered a black box, and the parameter variations are not visible to
 \synapse.

 A subset of tests have been performed with larger numbers of iterations (up to
 $10^9$), which confirmed the stability of results over that range.  Due to
 resource time limitations, we limit the results presented here to the range
 mentioned.

 \paragraph{Experiment Platform:} All profiling is performed on an off-the-shelf
 Intel Core i7 CPU (M620) with 4 cores, 8GB memory, Intel SSD 140GB (320-Series)
 under a Debian Linux with X86\_64 kernel v3.11.8-1.  Emulation experiments are
 performed on the same host, as well as on HPC resources, viz., Stampede, at the Texas Advanced Computing Center (TACC)~\cite{stampede_details},
 and on Archer, a Cray at EPCC~\cite{archer_details}.

 Stampede's compute nodes feature two 8-core Intel Xeon E5-2680 (Sandy
 Bridge) processors and an Intel Xeon Phi SE10P Coprocessor. The
 aggregate peak performance of the Xeon E5 processors is 2+PF -- we do
 not use the coprocessors in our experiments.  Each node has 32GB main
 memory, and a local 250GB HDD.  All I/O performed in our experiments
 is on that local hard drive.

 Archer is a Cray XC30 with two 12-core E5-2697 v2 (Ivy Bridge) series
 processors and 64GB main memory per node.  On Archer we also perform
 all disk I/O to /tmp, i.e.~to a local hard drive.

 The data sets produced and used in the experiments as presented below
 are freely available, as is the \synapse software itself; see the
 Software Availability paragraph.

 \setcounter{subsubsection}{0}
 \subsubsection*{Experiment 1 -- Profiling Self-Inference and Overhead}

\begin{figure}[t]
   \centering
   \includegraphics[width=0.50\textwidth]{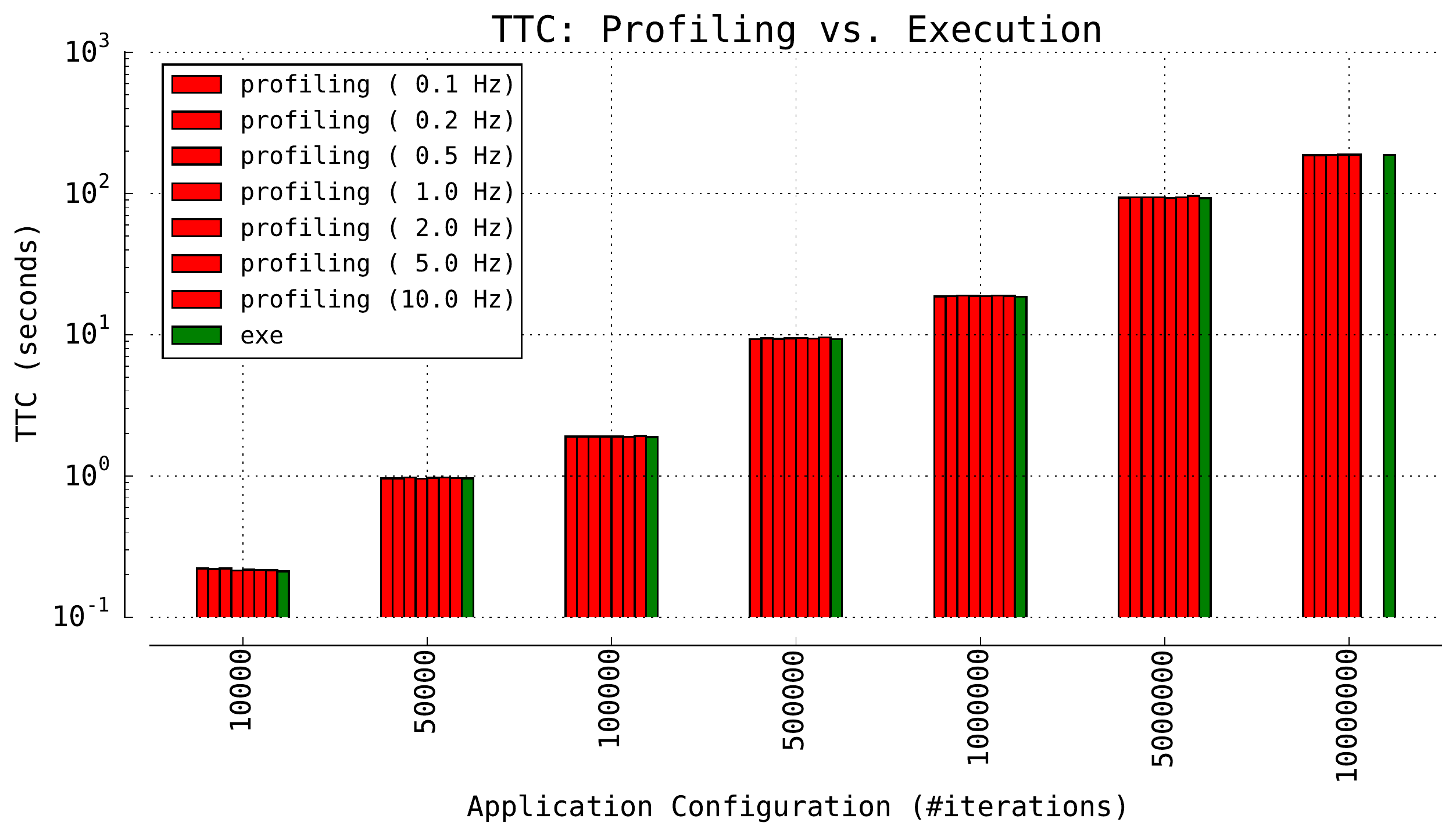}
   \caption{\textbf{Profiling Overhead:} While profiling does consume
       some additional resources, it does so in a way which does not
       impact the runtime of the profiles application.  The plots show
       constant runtime for all application configurations,
       independent of sampling rate. Note that the largest
       configuration misses one data sample due to limitations in the
       used database (see~\ref{sec:limits}).
     \label{fig:pro_overhead} 
   } 
   \end{figure}

 Figure~\ref{fig:pro_overhead} compares the TTC for two cases: pure
 application runs, and the execution of the application on the
 same resource under the \synapse profiler.  That measurement is shown
 for different application configurations (application runtimes) and
 different sampling rates.  The graph shows that the profiling overhead
 is negligible and remains so for the
 investigated range of problem sizes and sampling rates.

 \subsubsection*{Experiment 2 -- Profiling Consistency}

\begin{figure}[t]
   \centering
   \includegraphics[width=0.50\textwidth]{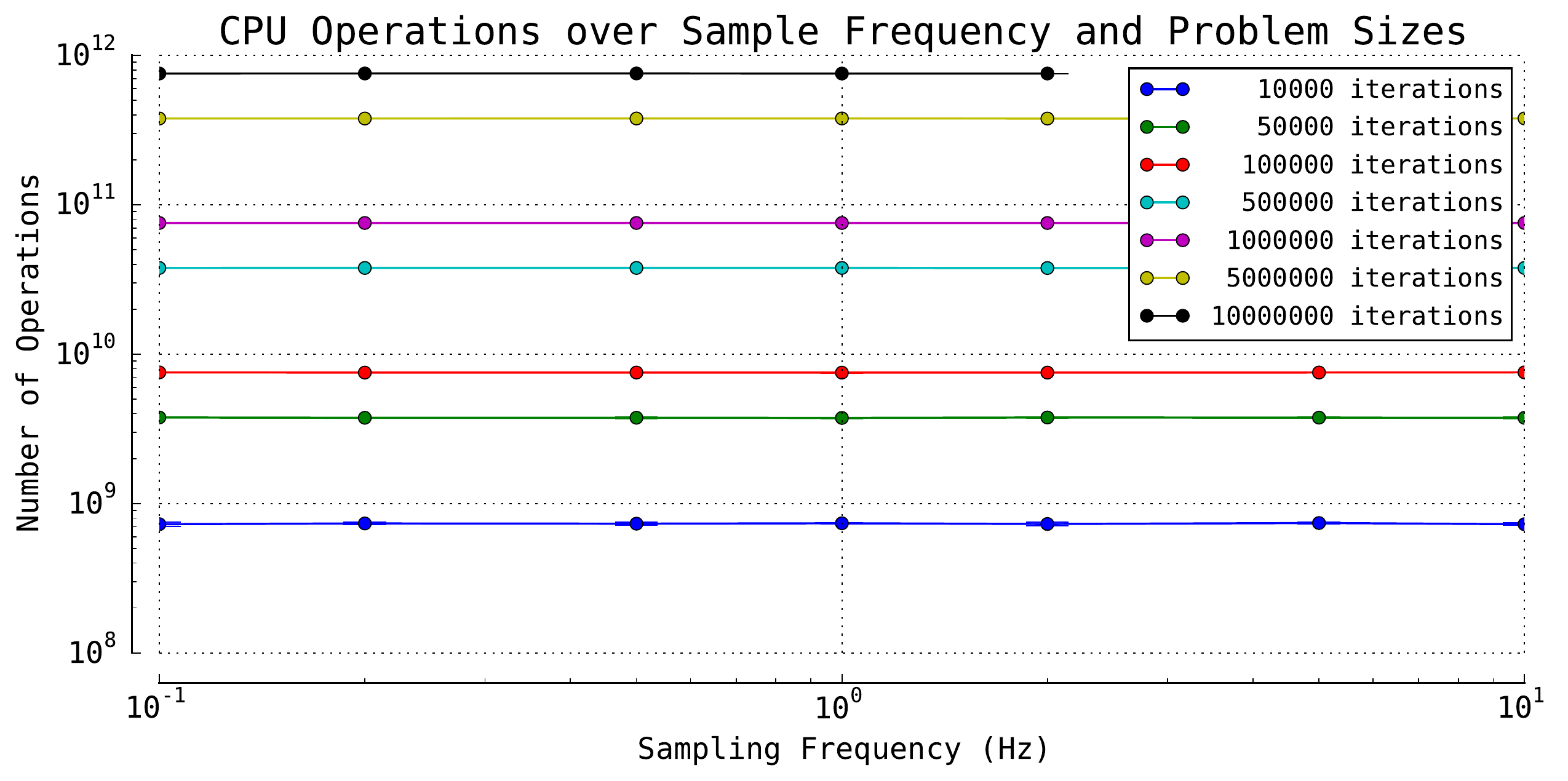}
   \caption{\textbf{Profiling Consistency:} Independent of the
       profiler sampling rate, \synapse reports very consistent
       values for consumed CPU operations, for a wide range of
       application runtimes (log/log scale, the plot includes
       error bars).
     \label{fig:pro_ops} 
   } 
   \end{figure}

\begin{figure}[t]
   \centering
   \includegraphics[width=0.50\textwidth]{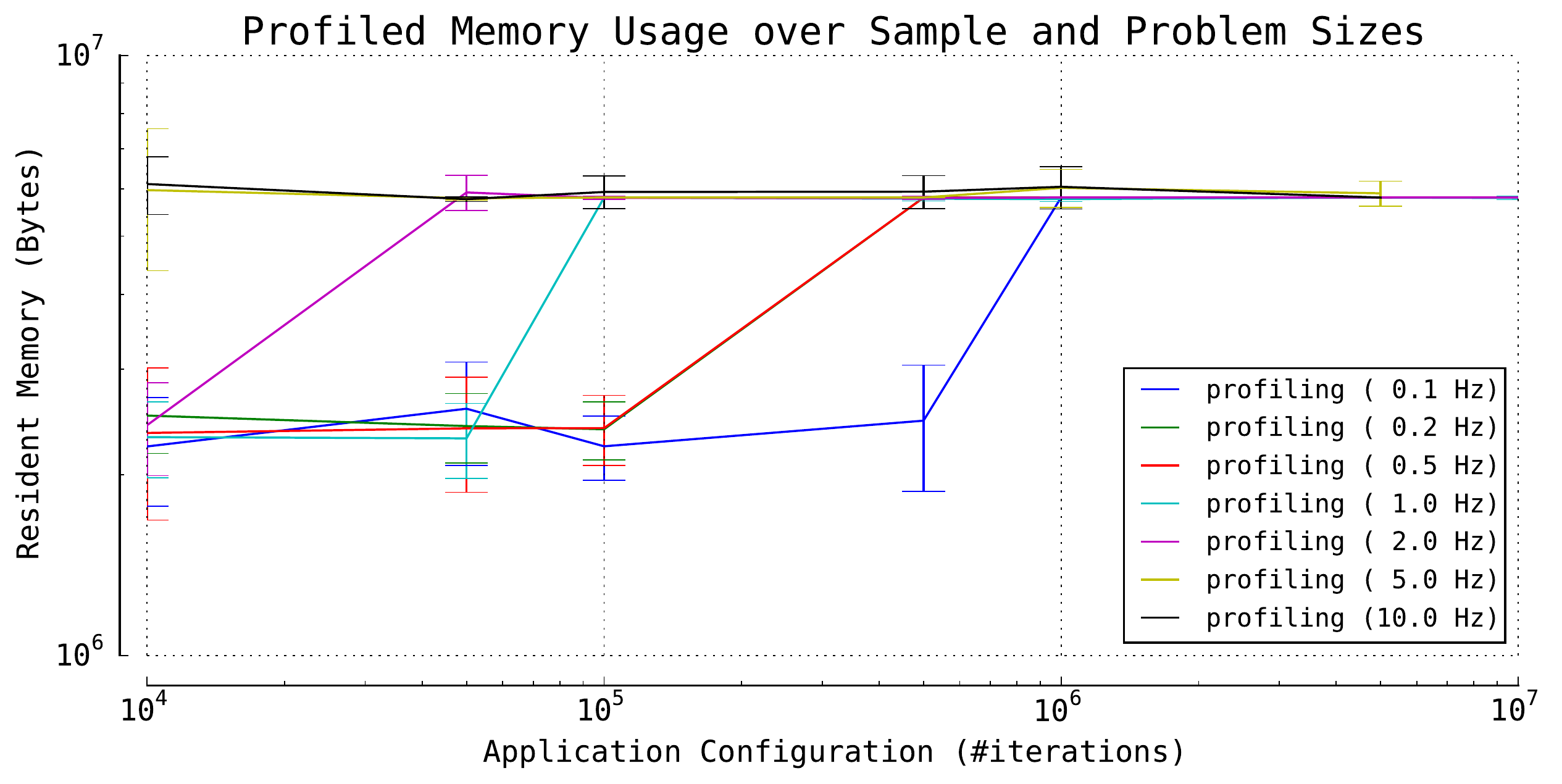}
   \caption{\textbf{Profiling Consistency:} For some metrics, the profiler
     requires sample rates to be smaller than application runtime.  For the
     example here (resident memory), the measure is underestimated by the
     profiler for sample rates that allow only one data sample to be taken over
     the course of the application runtime.  For multiple samples, the measures
     quickly stabilize.
     \label{fig:pro_mem} 
   } 
   \end{figure}

 We repeated profiling of the same application instances in the same
 environment.  While the non-zero standard deviation indicates some
 noise in the measured metrics, the distribution is in very good
 agreement with the distribution of the pure application TTC (see
 Figures~\ref{fig:pro_ops} and~\ref{fig:pro_mem}), which indicates the
 influence of system background.  The figure shows the profiling
 consistency over a range of application sizes and sampling rates.

 \subsubsection*{Experiment 3 -- Profiling as Emulation Input}

\begin{figure}[t]
   \centering
   \includegraphics[width=0.50\textwidth]{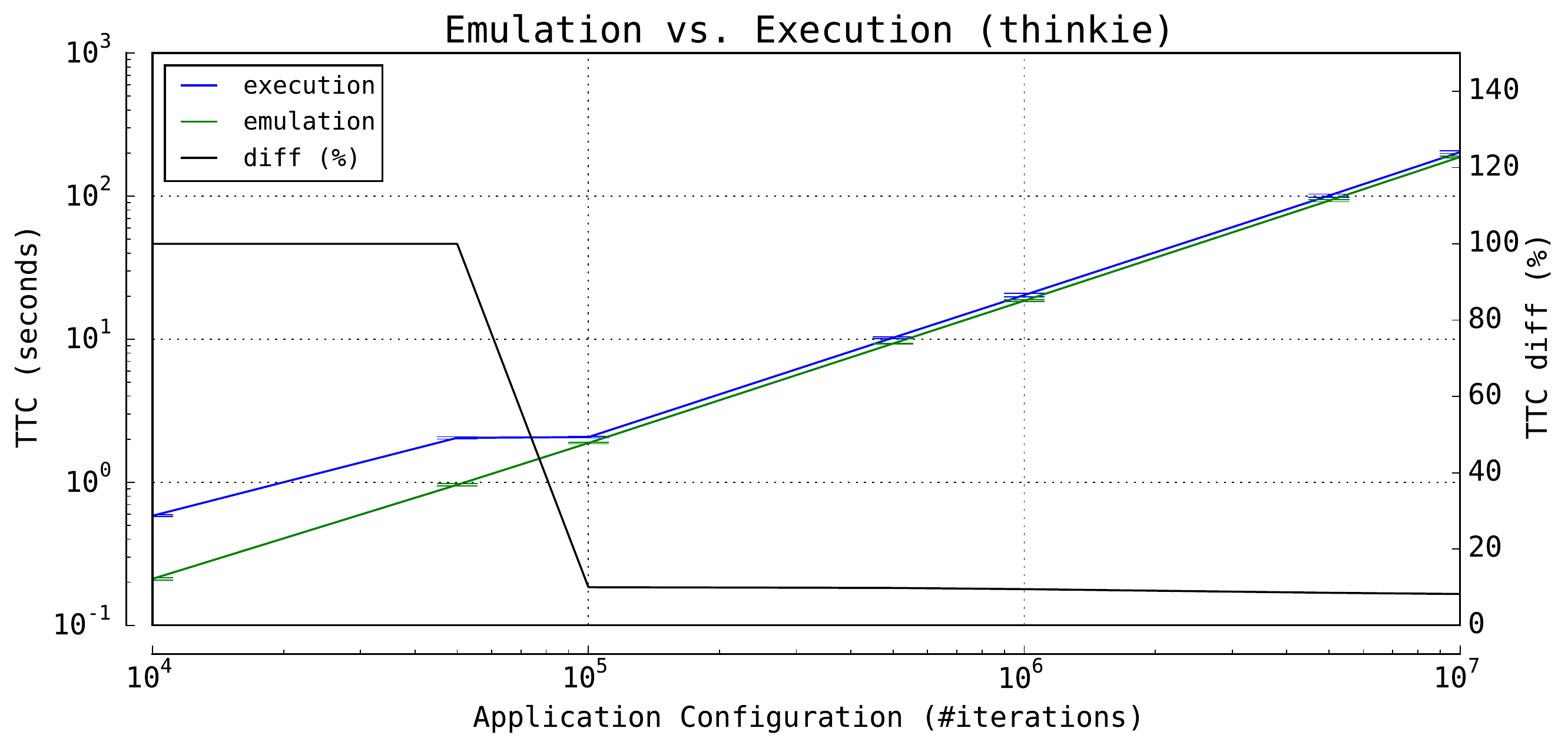}
   \caption{\textbf{Emulation Correctness:} When the application is emulated on
     the same host as used for profiling, the emulation represents the
     application characteristics excellently, as shown here for 7
     different application configurations (\#iterations).
     \label{fig:exe_emu_thinkie} 
   } 
   \end{figure}

 The ultimate purpose of \synapse's profiling is to feed \synapse's
 emulation.  Figure~\ref{fig:exe_emu_thinkie} compares the TTC of pure
 application execution versus emulated application runs, on the very
 same machine used for application profiling.  The graphs show that
 emulation tends to incur an overhead, specifically at startup time,
 which quickly becomes insignificant for applications running
 longer than a few seconds.

 As a self-check, we run the emulated application again under the
 profiler, and compared the reported resource consumptions: the values
 are in excellent agreement for any application instance running
 longer than a few seconds, as long as the sample rate is fast enough
 to result in at least two samples.  There exist some small deviations
 due to the memory footprint of the emulation driver (Python, C
 threads), but no other discernible difference.

 \subsubsection*{Experiment 4 -- Profiling as Portable Emulation Input}

\begin{figure}[t]
   \centering
   \includegraphics[width=0.50\textwidth]{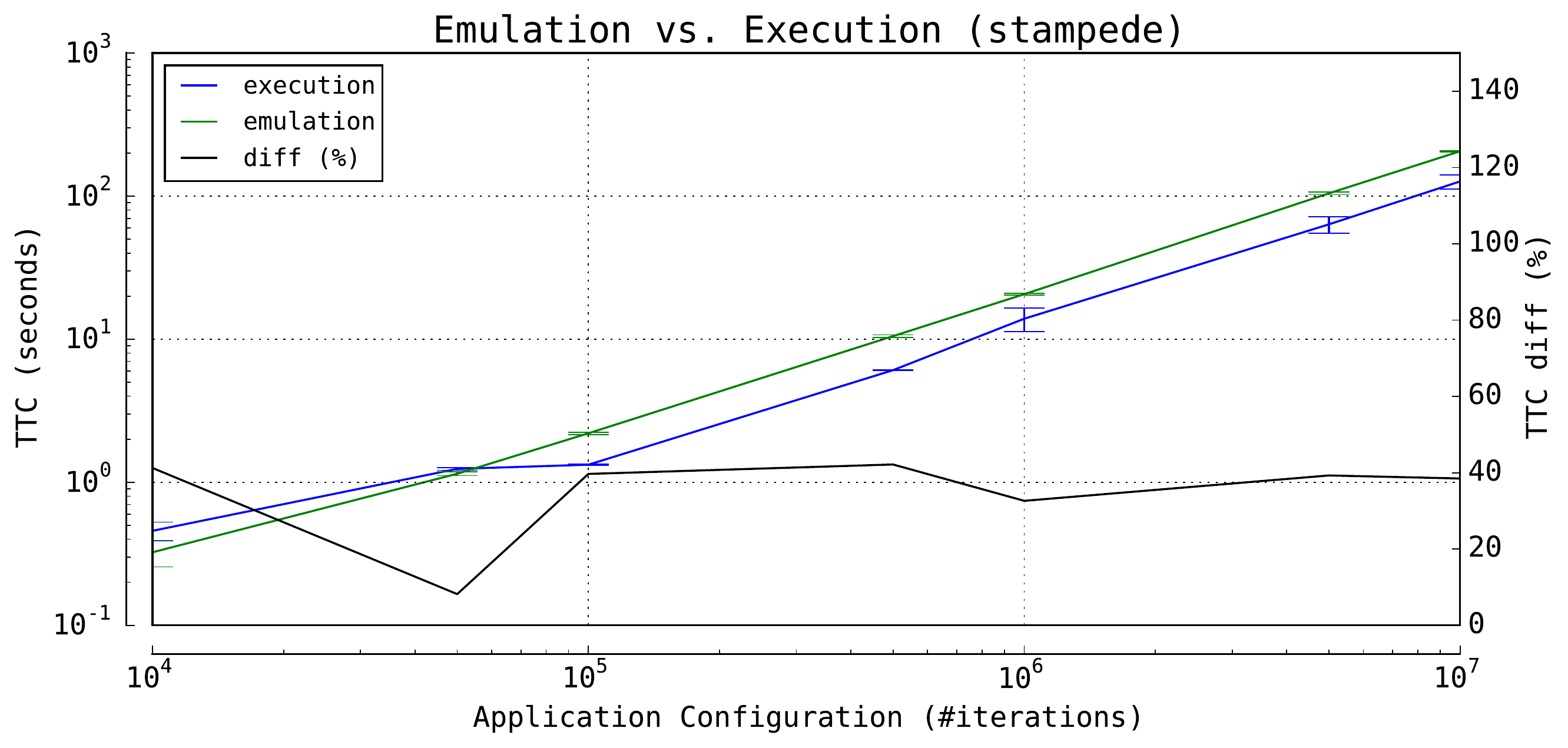}
   \caption{\textbf{Emulation Correctness:} While the emulation on Stampede
     (different to the machine on which profiling was done) is consistently
     faster compared to the application execution, it manages to capture the
     application's runtime trend precisely. The Y2 axis plots the \% difference
     between emulated time and actual execution time, which converges to about
     40\%.
    \label{fig:exe_emu_stampede} 
   } 
   \end{figure}

\begin{figure}[t]
   \centering
   \includegraphics[width=0.50\textwidth]{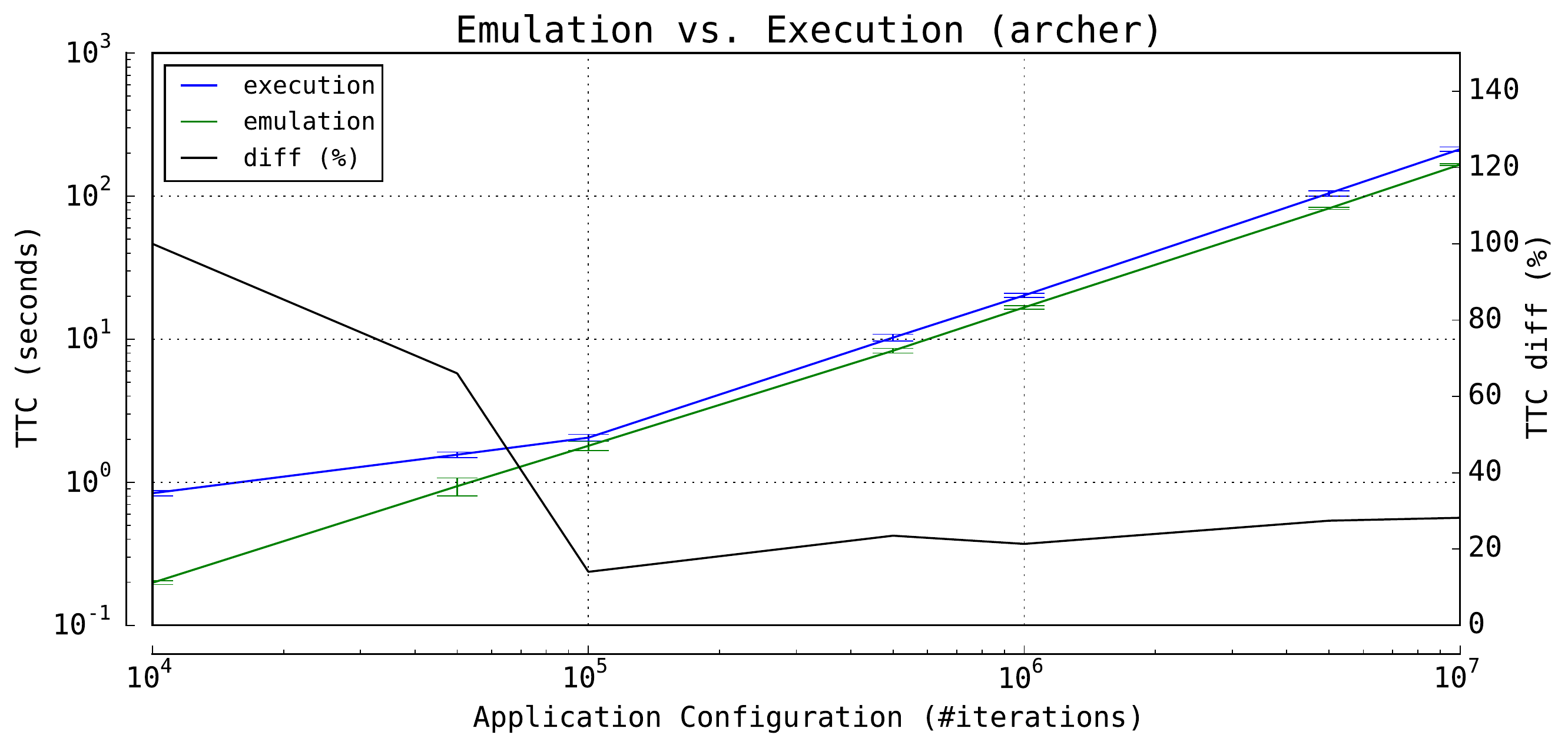}
   \caption{\textbf{Emulation Correctness:} The emulation runtimes on
       Archer are somewhat larger than actual execution times, but are
       still in reasonable agreement with the actual execution
       runtimes. The Y2 axis plots the \% difference between emulated
       time and actual execution time, which settles around around
       33\%.
       \label{fig:exe_emu_archer} 
   } 
   \end{figure}

 Figures~\ref{fig:exe_emu_stampede} and \ref{fig:exe_emu_archer}
 compare application execution and application emulation on resources
 \I{different} than the one used for application profiling,
 specifically on Stampede and Archer, respectively.  Again the TTCs
 for application execution and emulation resemble the application
 characteristics over a range of application sizes and sampling rates.

 The plots again show that the emulation overhead is significant for
 small application sizes, i.e., for application runtimes of about a
 second or smaller.  We find that acceptable, as our research focus
 lies on much longer living applications, and we assume that
 sub-second application runs will remain the exception in both
 high-throughput and high-performance distributed computing.

%
\section{Future Work} 
\label{sec:future}

 \subsubsection*{Sampling Rate}

 A high sampling rate has been shown to be able to capture application startup
 more accurately, and is necessary to handle short-running jobs.  At the same
 time, a high sampling rate incurs some emulation overhead.

 We will consider switching to an adaptive scheme which starts with a high
 sampling rate (10/sec), and after a few seconds, when we can expect to have
 captured the application startup, decrease the rate.  \synapse's code base does
 not assume a constant rate, but neither does it implement any sampling rate
 adaptation, yet.

 \subsubsection*{Networking, MPI}

 The most significant, but also most challenging next development step
 is the support for network profiling and emulation.  We consider it
 essential to capture the connectivity endpoints, and to attempt to
 perform actual data exchange to the remote network components, where
 possible.  That requires, however, changes to our current profiling
 approach, as a sample-based inspection seems insufficient to capture
 that information.  We consider to use |libc| call tracing for that
 purpose.

 A similar route seems useful to support the profiling and emulation
 of MPI and OpenMP applications.  A wide variety of MPI and OpenMP
 tracing tools and libraries exists, which we intent to investigate.

 \subsubsection*{Block-Level I/O Operations}

 The performance of disk I/O operations depends heavily on the
 storage system that is used, and on the granularity of the I/O requests toward
 that storage system.  \synapse currently captures neither of those,
 but we plan to use |blktrace| to obtain those information.  A
 prototype watcher plugin for |blktrace| exists.

 \subsubsection*{Resource Specificity}

 The experiments showed that \synapse application profiles are portable
 for emulation on other resources.  However, that is only valid when
 the application codes on the target resources are compiled with
 similar optimizations and against similar low-level numerical
 libraries.  
 
 We plan to investigate the \I{optional} introduction of an
 application-specific scaling factor that takes those differences into
 account.  Such a scaling factor could be determined via well defined
 probes that gauge the application performance toward a specific
 resource or resource configuration.  However, we expect the problem
 to persist in general, as it is unlikely that all configuration
 options can be determined automatically, or that any sample-based
 gauging will be representative for an application.  At the same time,
 the simplicity and resource-independence of \synapse is expected to
 remain an important design objective, even if that limits its
 emulation fidelity to some extent.

%
\section{Related Work}
\label{sec:related}


PAPI~\cite{papi2004} is widely used in the HPC community.  PAPI's sample based
evaluation of hardware counters is conceptually similar to \synapse profiling.
The simpler version in \synapse is based on standard Linux system utilities,
motivated by the use of resources where PAPI was not available and where we
lacked permissions to install it.  Also, using \texttt{perf} and other Linux
tools integrated better with elements of \synapse profiling, such as disk I/O or
memory allocation, which are not covered by PAPI.  There will likely be
convergence with PAPI for some of \synapse's profiling needs, so as to make the
\synapse profiling more portable and easier to maintain.

We are aware of only a few efforts to combine non-intrusive application
profiling with application emulation.  In~\cite{sodhi2004skeleton}, the authors
describe an approach to automatically derive application characteristics. It
focuses on tracing the application's communication calls, including MPI calls.
Other resource interactions are considered opaque and measured as times, and are
thus system dependent.  This approach works well for communication bound
applications.  The emulation represents a subset of the application: total
application TTCs are extrapolated under the assumption that the subset is
representative.

In ~\cite{skeleton-synapse15} Katz et al. work on a complementary approach of
\I{Application Skeletons}.  Skeletons do not include any mechanisms for
automatic application profiling, and thus require the user to specify resource
consumptions manually.  The focus of Skeletons is primarily on the
representation of logical and data dependencies between individual application
components: Application Skeletons can be used to represent a DAG of such
components.  Ref. \cite{skeleton-synapse15} discusses how \synapse can be used
to complement Application Skeletons, in that it provides configuration
parameters at the level of individual DAG components.

\amnote{we may or may not want to leave the paragraph below, depending on the
  intro...}  Note that a large body of work on the \I{simulation} of application
execution exists, which aims to predict application runtimes (and other metrics)
based on certain models of resources and runtime environments.  \Synapse is not
predictive, and thus does not relate to simulation approaches.

%
\section{Conclusions}
\label{sec:conclusions}

\synapse is capable of automatically deriving application characteristics, and
of configuring representative application emulation, for single threaded, scalar
applications.  While the application used to validate \synapse (Gromacs) is
representative of many other applications used in computational science, it
remains to be seen if this approach can suitably extend toward applications with
multiple threads or processes.

The profiling capability of \synapse has a low runtime overhead, and provides
stable, consistent results.  It requires no human intervention code
instrumentation, or exchange of libraries, and is fully transparent to the
application. It does however, need support at the system level and is
constrained to resources where |perf stat| can be executed by users. We believe
this not to be an issue in practice.

The emulation capabilities of \synapse provide a relatively accurate
representation of the application's behavior within constraints.  The main
contribution to emulation uncertainties arise in resource specific compile time
optimizations of the application codes, which are not covered when applying
application profiles across resources.  Nevertheless, \synapse's emulation
manages to capture the overall application characteristics and important trends
that determine TTC. When used on the same host as where profiling occurred,
\synapse provides high-fidelity emulation. Given the simplicity and low overhead
of usage we believe this will provide a useful contribution to the computational
science community. In fact, \synapse is used a proxy application for each of the
three use cases discussed in Section 2.

%

 \paragraph*{Software Availability}\label{sec:software}
 
 \synapse is available as Open Source Software, under the LGPL license,
 at~\cite{synapse_url}.  The experiments in this paper used version
 |v0.10|.  All scripts and configurations, along with the raw data
 sets and scripts for plotting, are available at~\cite{synapse_exp}.
 Please refer to the |README.md| file for instructions on how to
 reproduce the experiments.  Comments, feedback, and contributions to
 the software are welcome.  A bugtracker (which can also be used for
 feedback) is available at~\cite{synapse_tracker}.  When using this
 software, please reference~\cite{synapse}.


%
\section*{Acknowledgements}

\footnotesize{This work is supported by NSF ``CAREER'' ACI-1253644, NSF
  ACI-1440677 ``RADICAL-Cybertools'' and DOE Award DE-SC0008651. We acknowledge
  access to computational facilities on XSEDE resources via TG-MCB090174. We
  thank members of the RADICAL group, in particular Matteo Turilli for helpful
  discussions, suggestions \& testing, and Vivek Balasubramanian and Mark
  Santcroos for testing. We also thank Daniel S. Katz (U. Chicago) for useful
  discussions in the context of Skeletons and for improvements to the paper.}

%

\bibliographystyle{IEEEtran}
\bibliography{radical_publications,2015-wraps,enmd-local}

\end{document}

%% file: head.tex
\usepackage{booktabs}
\usepackage{graphicx}
\usepackage{amsmath}
\usepackage{amssymb}
\usepackage{color}
\usepackage{ifpdf}
\usepackage{float}
\usepackage[utf8]{inputenc}
\usepackage{multirow}
\usepackage{rotating}
\usepackage{subfig}
\usepackage{array}
\usepackage{mathtools}

\usepackage{moresize}
\usepackage{fancyvrb}
\usepackage{url}
\usepackage{booktabs}
\usepackage{listings}   
\usepackage{paralist}    
\usepackage{wrapfig}    
\usepackage{multirow}
\usepackage{ifpdf}
\usepackage{srcltx}
\usepackage{xspace}
\usepackage{keyval}  
\usepackage{color}
\usepackage{soul}

\definecolor{listinggray}{gray}{0.95}
\definecolor{darkgray}{gray}{0.7}
\definecolor{commentgreen}{rgb}{0, 0.4, 0}
\definecolor{darkblue}{rgb}{0, 0, 0.6}
\definecolor{purple}{rgb}{0.6, 0, 0.6}
\definecolor{middleblue}{rgb}{0, 0, 0.75}
\definecolor{darkred}{rgb}{0.4, 0, 0}
\definecolor{brown}{rgb}{0.5, 0.5, 0}
\definecolor{dkgreen}{rgb}{0,0.5,0}
\definecolor{orange}{rgb}{1,.5,0}
\definecolor{dandelion}{cmyk}{0,0.29,0.84,0}

\usepackage[normalem]{ulem}
\makeatletter
\def\cyanuwave{\bgroup \markoverwith{\lower3.5\p@\hbox{\sixly \textcolor{cyan}{\char58}}}\ULon}
\def\reduwave{\bgroup \markoverwith{\lower3.5\p@\hbox{\sixly \textcolor{red}{\char58}}}\ULon}
\def\blueuwave{\bgroup \markoverwith{\lower3.5\p@\hbox{\sixly \textcolor{blue}{\char58}}}\ULon}
\font\sixly=lasy6 
\makeatother